\documentstyle[11pt,aaspp4]{article}

\def\uv{{\it u,v} }
\newcommand{\etal}{{et al}\/.}

\lefthead{Looney \& Hardcastle}
\righthead{3C\,123}
\begin{document}

\title{Sub-Arcsecond Imaging of 3C\,123:\\
108-GHz Continuum Observations of the Radio Hotspots}
\author{Leslie W. Looney\altaffilmark{1}}
\affil{Max-Planck-Institut f\"{u}r Extraterrestrische Physik (MPE), Garching, Germany}

\and

\author{Martin J. Hardcastle\altaffilmark{2}}
\affil{H.\ H.\ Wills Physics Laboratory, University of Bristol, UK}

\altaffiltext{1}{Email: lwl@mpe.mpg.de}
\altaffiltext{2}{Email: m.hardcastle@bristol.ac.uk}

\begin{abstract}

We present the results of sub-arcsecond 108 GHz continuum
interferometric observations toward the radio luminous galaxy 3C\,123.
Using multi-array observations, we utilize the high \uv dynamic range
of the BIMA millimeter array to sample fully spatial scales ranging
from 0$\farcs$5 to 50$\arcsec$.  This allows us to make one-to-one
comparisons of millimeter-wavelength emission in the radio lobes and
hotspots to VLA centimeter observations at 1.4, 4.9, 8.4, and 15 GHz.
At 108 GHz, the bright, eastern double hotspot in the southern lobe is
resolved.  This is only the second time that a multiple hotspot region
has been resolved in the millimeter regime. We model the synchrotron
spectra of the hotspots and radio lobes using simple broken power-law
models with high energy cutoffs, and discuss the hotspot spectra and
their implications for models of multiple hotspot formation.

\end{abstract}

\keywords{galaxies: jets --- galaxies: structure --- radio continuum: galaxies}

\section{Introduction}

Hotspots in the jets of radio galaxies are manifestations of the
interaction between the jet and the intergalactic medium--- a strong
shock which converts some of the beam energy into relativistic particles
(\markcite{blan}Blandford \& Rees 1974).  Morphologically, hotspots
are bright compact regions toward the end of the jet lobe, primarily
observed in the radio with a few sources having optical counterparts
(e.g.\ \markcite{lat}L\"ateenm\"aki \& Valtaoja 1999).  First detected
in Cygnus A (\markcite{har}Hargrave \& Ryle 1974), hotspots are a
characteristic and ubiquitous feature in high luminosity, class FRII
radio galaxies (\markcite{fan}Fanaroff \& Riley 1974) that can provide
constraints on the energetics of the lobes and the powering of radio
loud active galactic nuclei.

Unfortunately, the simple, constant beam model of \markcite{blan}Blandford
\& Rees does not fully explain the common occurrence of multiple hotspot
regions in radio galaxies and quasars (cf. \markcite{laing}Laing 1989).
To accommodate these observations, two modifications have been proposed:
(1) the end of the beam precesses from point to point, the `dentist's
drill' model of \markcite{sche}Scheuer (1982) or (2) the shocked
material flows from the initial impact site to the secondary site,
the `splatter-spot' model of \markcite{will}Williams \& Gull (1985)
or the deflection model of \markcite{lons}Lonsdale \& Barthel (1986).
Both of these models predict that there should be a compact hotspot
at the jet termination; indeed, observations have shown that when the
jet is explicitly seen to terminate, it is always at the most compact
hotspot (\markcite{laing}Laing 1989; \markcite{lea1}Leahy et al. 1997;
\markcite{hard}Hardcastle \etal\ 1997).

However, the models in their simplest forms predict two essentially
different physical processes in the hotspots.  If the secondary (or less
compact) hotspots are the relics of primary (more compact) hotspots, as
suggested in the `dentist's drill' model, then the shock-driven particle
acceleration has ceased, and the spectrum of the continuum emission seen
toward these objects will steepen rapidly with increasing frequency as
a result of synchrotron aging and adiabatic expansion.  On the other
hand, the secondary hotspots in the `splatter-spot' or deflection models
still have ongoing particle acceleration as a result of outflow from
the primary hotspot, and as long as the observing frequency does not
correspond to an energy close to the expected high-energy cutoff in the
electron population, the spectral index will not be steeper than $\alpha$
= 1.0 (where S $\propto~\nu^{-\alpha}$), indicative of a balance between
spectral aging and particle acceleration.  Of course, it may be that
neither of these simple models can properly describe the physics of
the interaction.  For example, in a more sophisticated version of the
dentist's drill model (\markcite{cox}Cox, Gull, \& Scheuer 1991), the
disconnected jet material can continue to flow into the secondary hotspot,
causing particle acceleration for some time after the disconnection event.
This type of hybrid model will make predictions that will not always be
distinguishable from the simple cases.

Hotspots have been well studied with high resolution at radio
frequencies. To probe the hotspot regions at higher electron energies,
and test models for multiple hotspot formation, we present, in this
paper, the first high-resolution image of the FRII radio galaxy
3C\,123 in the 108 GHz continuum, focusing on the hotspot regions.
The radio galaxy 3C\,123 (z~=~0.218; \markcite{spin}Spinrad et
al. 1985) is one of the original FRII objects from
\markcite{fan}Fanaroff \& Riley (1974) and has an extremely high radio
luminosity, a highly unusual radio structure (\markcite{ril}Riley \&
Pooley 1978), and an optically peculiar host galaxy
(\markcite{hut1}Hutchings 1987; \markcite{hut2}Hutchings, Johnson \&
Pyke 1988).  With the highest resolution to date at these high
frequencies, we can compare the morphology and emission of the
hotspots to other high-resolution images at longer wavelengths.

Throughout the paper we use a cosmology with $H_0 = 50$ km s$^{-1}$
Mpc$^{-1}$ and $q_0 = 0$. With this cosmology, 1 arcsecond at the distance
of 3C\,123 corresponds to 4.74 kpc. The physical conditions we derive
in the components of 3C\,123 are not sensitive to the value of $H_0$.

\section{Observations and Imaging}

3C\,123 was observed in three configurations (C, B, and A) of the
9-element BIMA Array\footnote{The BIMA Array is operated by the
Berkeley Illinois Maryland Association under funding from the National
Science Foundation.} (\markcite{welch}Welch et al. 1996).  The
observations were acquired from 1996 November to 1997 February, with
the digital correlator configured with two 700~MHz bands centered at
106.04~GHz and 109.45~GHz.  The two continuum bands were checked for
consistency, then combined in the final images.  During all of 
the observations, the system temperatures ranged from 150-700 K (SSB).

In the compact C array (typical synthesized beam of
$\sim$8$\arcsec$), the shortest baselines were limited
by the antenna size of 6.1~m, yielding
a minimum projected baseline of 2.1~k$\lambda$ and good
sensitivity to structures as large as $\sim$50$\arcsec$.
This resolution is critical for obtaining an accurate
observation of the structure in the large-scale radio-lobes.
In the mid-sized B array (typical synthesized beam of
$\sim 2\arcsec$), the observations are sensitive to
structures as large as $\sim 10\arcsec$.
In the long-baseline A array (typical synthesized beam of
$\sim 0\farcs5$), the longest baselines were typically 
450~k$\lambda$.
With the high-resolution imaging of the hotspots, we can
make direct comparisons of the hotspots, and their components,
out to millimeter wavelengths.
The combination of the three arrays provide a 
well sampled \uv plane from 2.1~k$\lambda$ to 400~k$\lambda$.

The uncertainty in the amplitude calibration is 
estimated to be between 10\% and 15\%.
In the B and C arrays, the amplitude calibration was boot-strapped
from Mars.  In the A Array, amplitude calibration was done
by assuming the flux density of the quasar 3C\,273 to be 23.0 Jy.
This flux assumption was an interpolation through the 
A array configuration and supported by data from other observatories.
Absolute positions in our image have uncertainty due to the
uncertainty in the antenna locations and the statistical variation
from the signal-to-noise of the observation. These two
factors add in quadrature to give a typical absolute
positional uncertainty of $0\farcs$10 in the highest resolution image.

The A array observations required careful phase calibration. On long
baselines, the interferometer phase is very sensitive to atmospheric
fluctuations.  We employed rapid phase referencing; the observations
were switched between source and phase calibrator (separation of
9$\arcdeg$) on a two minute cycle, to follow the atmospheric phase
(\markcite{hold}Holdaway \& Owen 1995; \markcite{me}Looney, Mundy, \&
Welch 1997).  Since 3C\,123 was one of three sources included in the A
array calibration cycle, the time spent on-source was approximately 3
hours; thus, the noise in the high-resolution image is higher than would
otherwise be expected in a single track with the BIMA array.

\section{Results}

The data span \uv distances from 2.1k$\lambda$
to 430k$\lambda$, providing information on
the brightness distribution on spatial scales from
0$\farcs$4 to 60$\arcsec$.
In order to display the complete \uv information in the image plane, we 
imaged the emission with four different \uv
weighting schemes which include all of the \uv data and
stress structures on spatial scales of roughly 5$\arcsec$, 3$\arcsec$,
1$\arcsec$, and 0$\farcs5$.
These resolutions were obtained with natural weighting,
robust weighting (\markcite{dan}Briggs 1995) of 1.0, 
robust weighting of -0.2, and robust weighting of -0.6, respectively.
All data reduction was performed using MIRIAD (\markcite{bob}Sault, Teuben, \& Wright 1995), 
and the images shown were deconvolved using the CLEAN algorithm
(\markcite{hog}H\"ogbom 1974).

\begin{figure}
\includegraphics{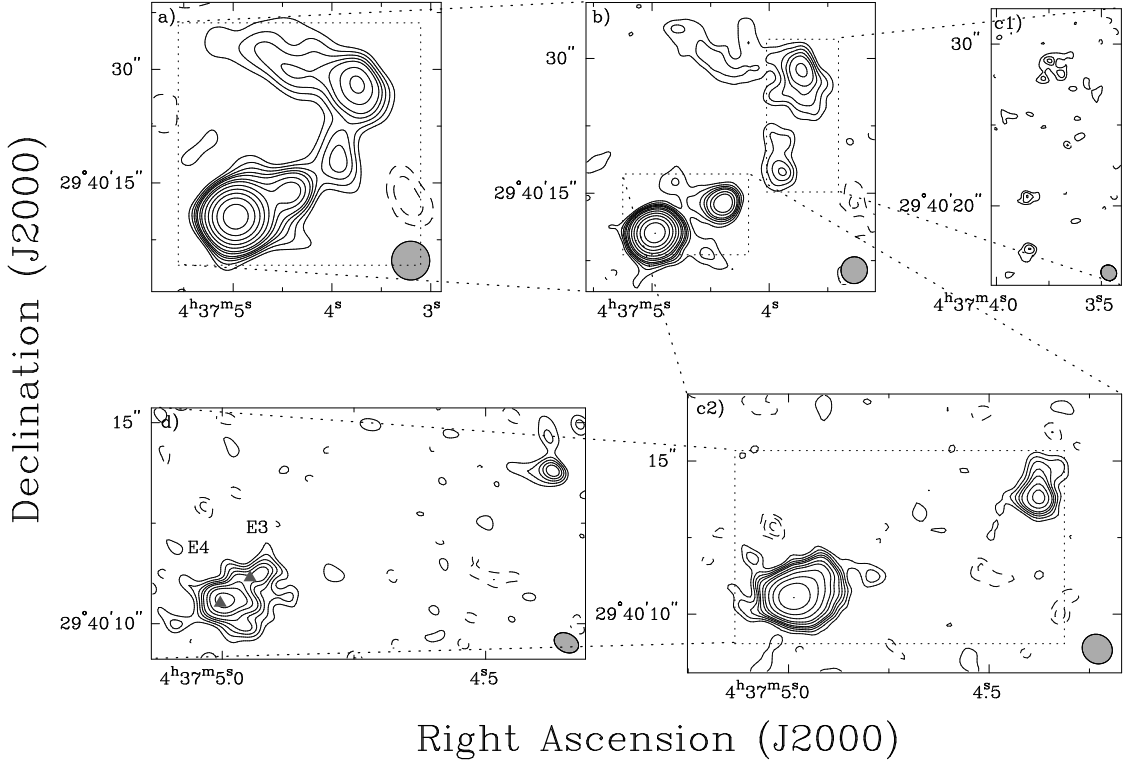}
\vspace{4.5in}
\caption{ The 108 GHz continuum emission from 3C\,123, imaged
with four different \uv weighting schemes.
All panels are contoured with steps of (-4 -3 -2 2
3 4 5 6 8 10 14.14 20 28.28 40 56.56 80 113.12) times the r.m.s.\
noise level $\sigma$ of each image.  (a) $\sigma$ = 3.9 mJy/beam and beam
of 5$\farcs$12 $\times$ 5$\farcs$03 P.A. = 4$\arcdeg$.  (b) $\sigma$ =
3.2 mJy/beam and beam of 2$\farcs$99 $\times$ 2$\farcs$93 P.A. =
-20$\arcdeg$.  (c) c1 and c2 are sections of the same image.  c1 is a
close-up of the northwest region; there are three 4$\sigma$ peaks.  c2
is a close-up of the southeast region.  $\sigma$ = 3.5 mJy/beam and beam
of 1$\farcs$02 $\times$ 0$\farcs$93 P.A. = 48$\arcdeg$.  (d) The two
grey triangles mark the centimeter positions of the two main
components of the eastern hotspot (E3 and E4) from Hardcastle et al
(1997).  $\sigma$ = 4.5 mJy/beam and beam of 0$\farcs$64 $\times$
0$\farcs$46 P.A. = 62$\arcdeg$.  }
\label{maps}
\end{figure}

The 108 GHz continuum emission from 3C\,123, imaged at the four resolutions
mentioned above, is shown in Fig.\ \ref{maps}.  In this Figure, each successive
panel is a higher-resolution zoom, beginning with the 5$\arcsec$ image.
Fig.\ \ref{maps}a shows the large-scale overall jet-lobe structure, which is
very similar to lower frequency images (e.g. \markcite{hard}Hardcastle
et al. 1997) and other low resolution millimeter images at 98 GHz
(\markcite{ok}Okayasu, Ishiguro, \& Tabara 1992).  Our observations, which
have more sensitivity to large-scale structure and better signal-to-noise
than the 98 GHz data, do not detect the extended emission to the south of
the bright eastern hotspot that is seen at longer wavelengths (component
F of \markcite{ril}Riley \& Pooley 1978). We also do not detect feature
H of \markcite{ok}Okayasu \etal\ (1992), which does not in any case
correspond to any feature seen on lower-frequency radio images.

In Fig.\ \ref{maps}b, the 4 major sources of millimeter emission at
3$\arcsec$ resolution are clearly distinguished-- from east to west,
the eastern hotspot, the core, the western hotspot, and the northwest
lobe, respectively.  As the resolution increases to $\sim$1$\arcsec$
in Fig.\ \ref{maps}c1 and \ref{maps}c2, the western hotspot and the
northwest lobe corner are resolved into three peaks that contain only
a small fraction of the large scale flux.  Since the interferometer is
acting as a spatial filter, this implies that the northern lobe
consists mainly of large-scale emission; however, the eastern hotspot
is dominated by compact emission at this resolution.  In the highest
resolution image, Fig.\ \ref{maps}d, the eastern hotspot is resolved
at a principal axis of $\sim$38$\arcdeg$, while the core is a point
source. The western hotspot is too faint to be seen in this image.
Our image of the eastern hotspot looks very similar to high resolution
8.4 GHz observations (\markcite{hard}Hardcastle \etal\ 1997), which
resolve the hotspot into two components --- an extended southeastern
component (E4), which corresponds to the peak of the 108 GHz image,
and a very compact northwestern component (E3), which accounts for the
extension seen in the present image.

\section{VLA data and Spectral Indices}

To compare our data with observations at longer wavelengths, we
obtained existing Very Large Array (VLA) data or images at 1.4, 5, 8.4
and 15 GHz. The 1.4 GHz image was taken from \markcite{lea2}Leahy,
Bridle \& Strom (1998) based on observations with the VLA A
configuration, the 5 and 15 GHz were re-reduced observations by
R. A. Laing from the VLA archive using A and B configurations and B
and C configurations respectively, and the 8.4 GHz data were from
\markcite{hard}Hardcastle \etal\ (1997), using A, B and C
configurations. All these datasets have shortest baselines very
similar to that of our BIMA data, so that they sample comparable
largest angular scales; with the exception of the 1.4 GHz data, they
are also comparable in longest baseline and thus angular
resolution. Flux density scales were calibrated using observations of
3C\,48 and 3C\,286; we applied a correction to the flux levels of the
15 GHz B-configuration data to compensate for an estimated 7\%
decrease in the flux density of 3C\,48 between the epoch of
observation (1982 August 06) and the epoch at which the flux density
coefficients for 3C\,48 used in AIPS were measured (1995.2).

\begin{figure}
\plotone{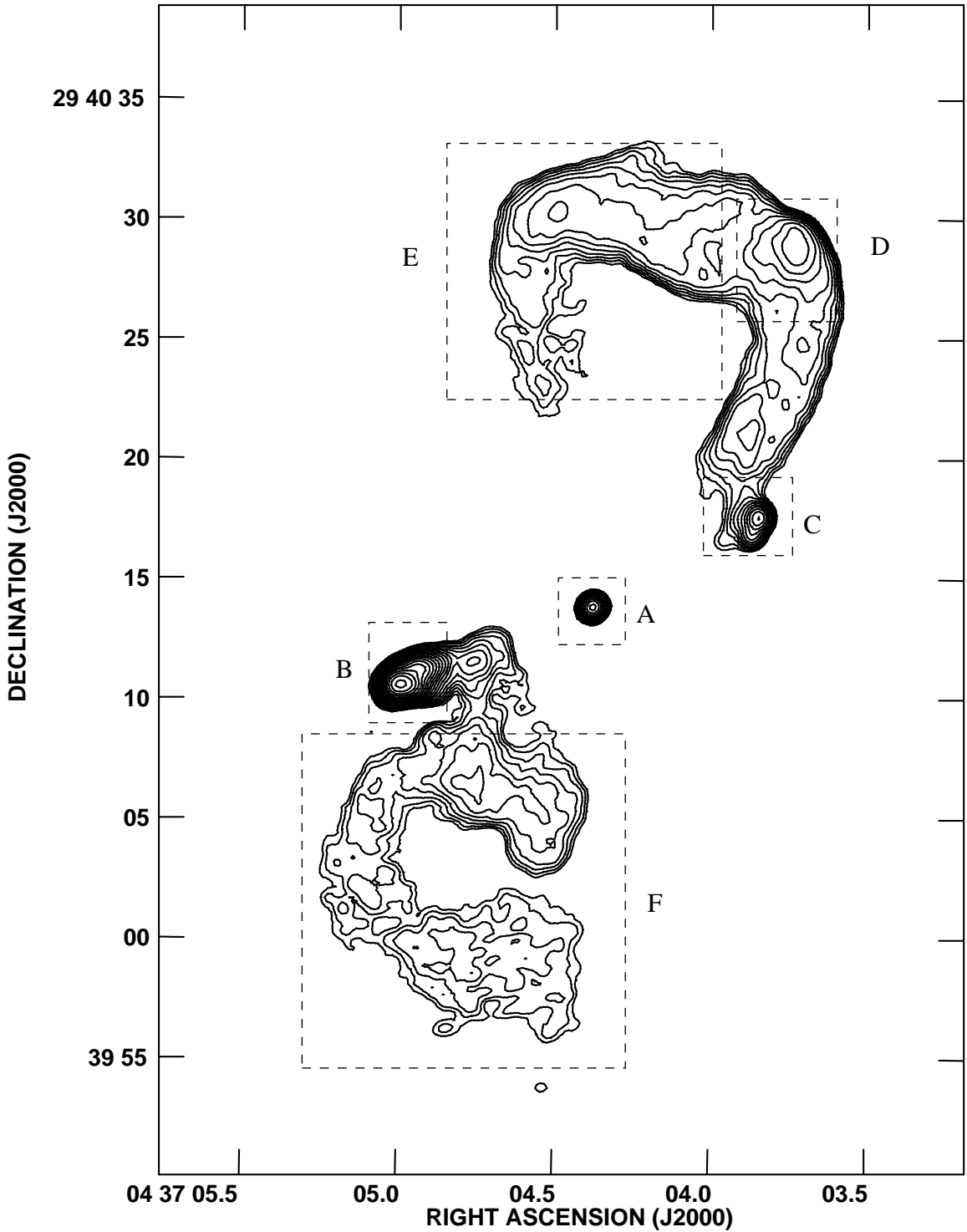}
\caption{A VLA contour map of 3C\,123 at 8.4 GHz. The resolution of this
map is $0\farcs 6$ and the contour levels are 4 mJy beam$^{-1}$
$\times (1, \sqrt{2}, 2, 2\sqrt{2}, 4, \dots)$ (there are no
equivalent negative contours). The approximate
locations of the regions in which flux density measurements were made are
shown as dashed boxes, labeled with letters as follows: A, core; B, E
hotspot; C, W hotspot; D, NW corner; E, N lobe; F, S lobe. As
discussed in the text, the actual regions used were polygonal areas
defined with MIRIAD.}
\label{regions}
\end{figure}

Having imaged the VLA data, we measured the flux densities of the
various components of 3C\,123 using the regions specified on the 8.4
GHz VLA image in Fig.\ \ref{regions}.  These flux densities are
tabulated in Table \ref{fluxes}.  Except where otherwise stated in the
final column of the Table, they are derived by integration using
MIRIAD, from aligned images, convolved to the same (3$\arcsec $)
resolution, with polygonal regions defined on low-frequency
images. This process ensures that we are measuring the same region at
each frequency. The exceptions are the flux density of the core, which
was measured by fitting a Gaussian to the matched-resolution maps, and
the flux densities of the two components of the E hotspot, which were
measured from maps with resolution matched to the highest resolution
of the BIMA data. Using these flux densities, we derived a spectral
index between each of the 5 frequencies (4 two-point spectral
indices).  Table \ref{spices} lists these spectral indices for each
component in the four bands.

The radio core shows an approximately flat spectral index across the
radio and millimeter bands. The 8.4 GHz data were taken in 1993--1995
while the other radio frequencies were taken in 1982--1983, so we are
comparing data separated in time by a decade, but there was no
evidence for core variability on timescales of years in the
observations at different epochs that make up the 5, 8.4 and 15 GHz
datasets, and the similarity in the flux densities at 8.4 GHz and 5
and 15 GHz [cf. also the 15 GHz core flux density of 120 mJy from
\markcite{ril}Riley \& Pooley (1978) and the 5 GHz core flux density
of 99 mJy measured from the MERLIN images of \markcite{hard}Hardcastle
\etal\ (1997)] suggests that there is little variability even on
timescales of decades at centimeter wavelengths, contrasting with the
variability found in some other well-observed radio galaxies with
bright radio cores. However, our 108 GHz core flux density is a factor
3 lower than the flux density measured by \markcite{ok}Okayasu \etal\
(1992) between 1989 and 1990 at 98 GHz. Either the spectrum cuts off
very sharply between these frequencies --- more sharply than would be
expected in a synchrotron model --- or, more probably, the core is
more variable at higher frequency. It is generally found in studies of
core-dominated objects that the amplitude of nuclear variability is
higher in the millimeter band than at centimeter wavelengths, a fact
which can be explained in terms of synchrotron self-absorption effects
at lower frequencies (e.g., \markcite{hugh}Hughes, Aller \& Aller
1989). Unfortunately, little is known about the millimeter-wave
variability of lobe-dominated objects like 3C\,123.

All the other components of the radio source have relatively steep spectra
even at centimeter wavelengths. As expected, the flattest spectra are
observed in the hot spots. We cannot distinguish between the NW and SE
component of the E hotspot, within the errors, on the basis of their
high-frequency spectral indices, and the W hotspot, also detected at
108 GHz, has a comparable spectrum. The southern lobe (all extended
emission to the south of the eastern hotspot, see Fig.\ \ref{regions})
has spectral indices which indicate a spectral cutoff at centimeter
wavelengths, so it is not surprising that we do not detect it at 108
GHz. However, the northern lobe (the extended emission E and S of the
`NW corner') shows no strong indication of a spectral cutoff even at
millimeter wavelengths. 

\section{Spectral Fitting}

In order to investigate the synchrotron emission, we fit simulated
spectra to the different components of the source, using the code from
\markcite{hard2}Hardcastle, Birkinshaw \& Worrall (1998).  We assume
an injection energy index for the electrons of 2, corresponding
to a low-frequency spectral index of 0.5, since we cannot derive an
injection index from any of our existing data; the 81.5 MHz scintillation
measurements of \markcite{read}Readhead \& Hewish (1974) suggest a flatter
spectral index for the hotspots, but this low frequency may be below
a spectral turnover due to synchrotron self-absorption or a low-energy
cutoff in the electron energy spectrum, as seen in the hotspots of Cygnus
A (\markcite{chris}Carilli \etal\ 1991). To find magnetic field strengths,
we assume equipartition between the electrons and magnetic fields, with
no contribution to the energy density from relativistic protons. The
choice of an equipartition field does not affect our conclusions about
spectral shape, but does affect our estimates of break and cutoff electron
energies. Since the fitting is essentially done in the frequency domain,
all energies quoted may be scaled by a factor $\sqrt{B_{\rm eq}/B}$ if
the field deviates from equipartition. We perform $\chi^2$ fitting of the
simulated spectra by combining the systematic errors in flux calibration
(fixed at 2\% for the VLA data and 10\% for the BIMA data) with the
statistical errors tabulated in Table \ref{fluxes}; the systematic
errors are the dominant source of error for the VLA data. Because the
systematic errors are uncorrelated from frequency to frequency, this
procedure is valid when fitting spectra, though not when {\it comparing}
fluxes or spectral indices from different parts of the source.

We consider two basic models for the electron energy spectrum. Both
have high-energy cutoffs, but one has a constant electron energy index
of 2, while the other is a broken power law model, allowed to steepen
from an electron energy index of 2 to 3 at a given energy.  The latter
is appropriate for a situation in which particle acceleration is being
balanced by synchrotron losses or in which loss processes are
important within the hotspot (\markcite{pach}Pacholczyk 1970;
\markcite{heav}Heavens \& Meisenheimer 1987). These two models are
equivalent to models (i) and (ii) of \markcite{meis}Meisenheimer
\etal\ (1989), respectively.  

\subsection{Component Fitting Results}

The fitting results are tabulated in Table \ref{fits}.  We find that
model (i), the simple, single power-law, never fits the data well, and
that in half of the component fits, model (ii), the broken power-law
model, fits well with a very high-energy cutoff (labeled as ``break''
in Table \ref{fits}).  For the rest of the components, a broken-power
law and a high-energy cutoff within our data's frequency range is
necessary (labeled as ``both'' in Table \ref{fits}). For the chosen
model, we tabulate the equipartition magnetic field strength in nT and
the best-fitting break energies and, where appropriate, cutoff
energies in GeV.

\begin{figure}
\plotone{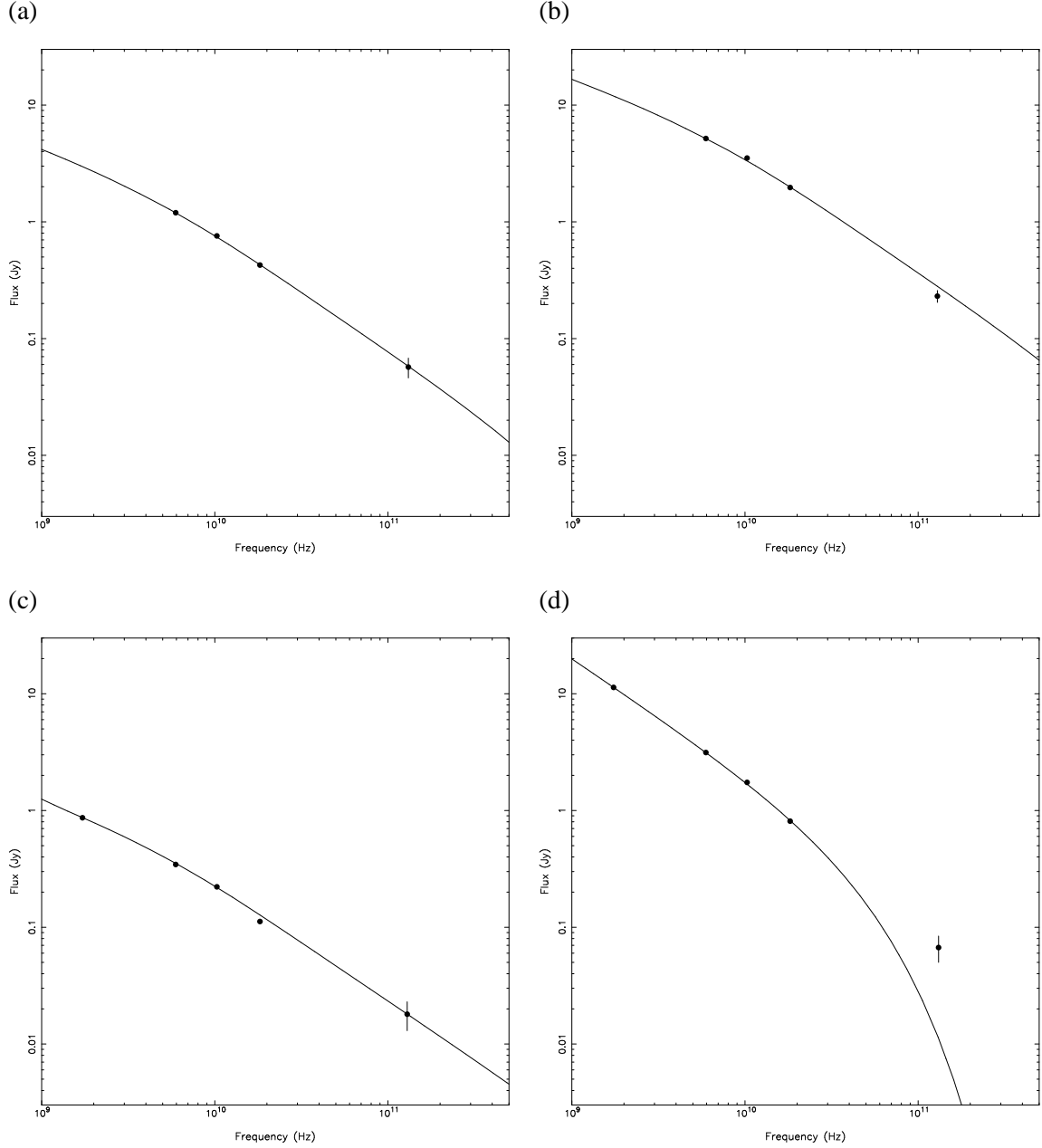}
\caption{Flux densities and best-fitting synchrotron spectra for
components of 3C\,123. (a) Northern component of E hotspot (E3) (b)
Southern component of E hotspot (E4) (c) Western hotspot (d) North
lobe. The flux densities are taken from Table \ref{fluxes}; the
synchrotron spectra plotted (as solid lines) are those tabulated in Table
\ref{fits}. Note that frequencies are plotted in the source frame.}
\label{spectra}
\end{figure}

The NW component of the E hotspot (E3) is well fit with the
break model (Fig.\ \ref{spectra}a), but it is very poorly fit with 
break models having a energy cutoff within our data frequency range;
all of the best high cutoff fits to our data have cutoff energies 
above $10^{10}$ eV (corresponding to $>200$ GHz). 
This is due mainly to the essentially constant spectral index 
between 8 and 108 GHz.
The SE component of the E hotspot (E4) is also best fit with
a broken power-law spectrum, although not as well, and only poorly
with a high-energy cutoff spectrum (Fig.\ \ref{spectra}b).

These results differ from the conclusion of \markcite{meis2}Meisenheimer,
Yates \& R\"oser (1997), who prefer a model with only a high-energy cutoff
as a fit to the overall spectrum of the eastern hotspot.  This may be the
result of subtle measurement differences in the regions and frequencies
used by \markcite{meis2}Meisenheimer \etal , who took flux densities for the
E hotspot from a variety of sources in the literature, or it may be
the effect of combining the two hotspot regions.  Our results are more
consistent with the model favored by \markcite{meis}Meisenheimer \etal\
(1989).

The W hotspot is also best fit with a broken power-law model
(Fig.~\ref{spectra}c), although no fit is particularly good because of
the anomalously flattening spectral index between 15 and 108 GHz that
our simple models cannot reproduce. The effect may be due to a bad
data point at 15 GHz, but it should be noted that we are not resolving
the two components of this hotspot (\markcite{hard}Hardcastle \etal\
1997), so the spectral situation is probably more complex than is
represented by our simple one-component model. Again, a high-energy
cutoff within our frequency range fits the data even more poorly.

Although the northwest corner region is resolved out at high
resolution, it dominates the western side of our low-resolution 108
GHz images (Fig.~\ref{maps}a).  The spectrum of this region is
smoothly curved from centimeter to millimeter wavelengths. It is
poorly fit with a single power-law and cutoff model, but reasonably
well fit with a spectral break model; however, better fits are
obtained with a model with a high-energy cutoff as well as a spectral
break (though the improvement is not significant on an F-test) because
of the steep 15--108 GHz spectral index.

The northern lobe's spectrum is poorly fit with the break model or
with a high-frequency cutoff; even the combination of the two, though
a substantial improvement, gives a clearly poor fit, modeling the 108
GHz data badly (Fig.\ \ref{spectra}d), because of the way the spectrum
first curves between 8 and 15 GHz, then remains straight between 15
and 108 GHz (Table \ref{spices}). A \markcite{jaf}Jaffe \& Perola
(1973) aged synchrotron spectrum is also a poor fit, though it does
represent the 108 GHz data better.  Like the northern lobe, the
southern lobe is best fit with a spectral break and high-energy
cutoff, but again the fits are not particularly good.

Overall, the regions required models with broken power-laws to achieve good
fits, but the three hotspot component models have high-energy cutoffs
significantly above 108 GHz, while the three other regions required
energy cutoffs within our data frequency range.

\subsection{Spectral Model Interpretation}

The mm-to-cm spectra of both components of the eastern hotspot, resolved
at millimeter wavelengths for the first time in our observations, are
consistent with a simple, spectral break model, as expected for regions
in which ongoing particle acceleration is balanced by synchrotron
losses. There is no evidence for significant spectral differences
between the two hotspot components, which implies either that particle
acceleration (and hence energy supply) is still ongoing in the less
compact SE component, as in the model of \markcite{will}Williams \&
Gull (1985), or that it was disconnected from the energy flow less
than $\sim 1.5 \times 10^4$ years ago, assuming \markcite{jaf}Jaffe \&
Perola (1973) spectral aging on top of the broken power-law model for
the electron spectrum and an aging field equal to the equipartition field
in Table \ref{fits}. (The estimate of $1.5 \times 10^4$ years is a 99\%
confidence limit with $\Delta \chi^2 = 6.6$. We neglect the possible
effects of adiabatic expansion.)

The three non-hotspot regions studied all show evidence for a
high-energy cutoff in addition to the broken power-law spectrum of the
hotspots. However, it is clearly more difficult to draw conclusions from the
fitted spectra. The fact that the fitted break energies in the lobes
are much lower than the break energies in the hotspots may suggest
that the assumption of equipartition is wrong in one or both regions,
with $B$-field strengths deviating from their equipartition values by
up to a factor $\sim 40$. However, X-ray observations suggest that
both in the hotspots and in the lobes of other radio galaxies the
magnetic field strength is close to equipartition with the energy
density in relativistic electrons (\markcite{harris}Harris, Carilli \&
Perley 1994; \markcite{fei}Feigelson \etal\ 1995;
\markcite{tsak}Tsakiris \etal\ 1996). (The equipartition assumption in
the hotspots of 3C\,123 will be tested by forthcoming {\it Chandra}
observations.) Instead, the lower break energies seen in the lobes may
simply be due to adiabatic expansion of the electron population as it
leaves the hotspot. Radial expansion by a factor $\epsilon$ moves the
electron energy spectrum down by a factor $\epsilon^{-1}$, so the
estimated change in break energies between the hotspots and lobes
implies expansion out of the hotspots by factors up to $\sim 6$,
though we emphasize that the break energies in the lobes are only
weakly constrained by the data. These factors are rather higher than
those that would be estimated from the ratio of magnetic fields
between lobe and hotspot (field strength $B \propto \epsilon^{-2}$ on
adiabatic expansion). If either expansion has taken place or the
magnetic field in the lobes is much weaker than equipartition, the
high-energy cutoffs fitted to the lobe data cannot be said to be
unambiguously due to spectral aging; to take the most extreme example,
shifting the break energy of the S lobe up to match that of the E4
component of the E hotspot brings the corresponding cutoff energy up
to 30 GeV, which is not ruled out by our data.  In any case, the
expected aged spectrum depends on the detailed order of expansion and
aging, and the lobes are probably not spectrally homogeneous, so we do
not attempt to fit aging models to the data.

Unlike the lobe spectra, the best-fit spectrum of the NW corner shows
a break energy that is comparable to those in the W hotspot, and is
certainly consistent within the large errors introduced by
uncertainties in the geometry and field strength. The brightening here
may be due either to particle reacceleration in this region or simply
to compression. The fact that the break energy is higher than that in
the W hotspot while the magnetic field strength is lower might seem to
favor a reacceleration model, but if, as seems likely, the hotspots
are transient features, the present-day properties of the W hotspot do
not necessarily reflect those of the hotspot that was present when the
material now at the NW corner was first accelerated. The same caveat,
of course, applies to a comparison of the hotspot and lobe spectra.

\section{Conclusions}

We have presented the first sub-arcsecond millimeter wavelength
continuum imaging of the radio galaxy 3C\,123, resolving the eastern
hotspot.  These are only the second observations at millimeter
wavelengths to resolve a double hotspot pair. Hat Creek and later BIMA
observations of the bright, nearby classical double radio galaxy
Cygnus A (\markcite{mel1}Wright \& Birkinshaw 1983;
\markcite{mel3}Wright \& Sault 1993; \markcite{mel2}Wright, Chernin \&
Forster 1997) resolve both the eastern and western double hotspots in
that source, and, as in the case of 3C\,123, it is found that in
Cygnus A there is little or no clear spectral difference between the
primary (more compact) and secondary (more diffuse) hotspots. Thus, in
both these sources it is impossible to say whether or not there is
continued energy supply to the secondary hotspot. The short
synchrotron lifetimes at millimeter wavelengths mean that if the
secondary hotspots {\it are} disconnected from the energy supply, as
in the `dentist's drill' model, the disconnection must have taken
place on timescales which are much shorter (by factors of $>100$) than
the lifetime of a typical radio source.  Indeed, numerical simulations
suggest that such short-timescale transient hotspot structures are
expected in low-density radio sources (\markcite{norm}Norman 1996).

In both 3C\,123 and Cygnus A, there is no clear evidence in the radio
structure for continuing outflow between the primary and secondary
hotspotsi.  Specifically, there are no filaments connecting the
eastern hotspots in 3C\,123 together, as there are in several other
multiple-hotspot sources or even in the western hotspot pair of 3C\,123
(\markcite{hard}Hardcastle \etal\ 1997), and the suggestion
that the hotspots in Cygnus A are connected by an outflow marked by a
ridge seen in the radio is inconsistent with the pressure gradients in
the lobes, as pointed out by \markcite{cox}Cox \etal\ (1991).  Overall,
therefore, the situation in these two sources seems most consistent with
the picture of \markcite{cox}Cox \etal , in which the bright secondary
hotspots are recently disconnected remnants of earlier primaries and are
still being, or have been until recently, powered by continued inflow
of disconnected jet material.  These models predict that sources should
exist in which the secondary hotspots are genuinely no longer powered,
as in the original dentist's drill model; such sources {\it should},
observed at the right time in the evolution of their hotspots, show a
clear spectral difference between the primary and secondary hotspots at
millimeter wavelengths. To find them, it seems likely that it will be
necessary to look at sources with more typical double hotspot structure
and without the dominant, compact secondary hotspots of 3C\,123 and
Cygnus A; we have BIMA data for such a source (3C\,20) and will report
on our results in a future paper.

On larger scales, our observations of 3C\,123 show a striking
difference in the spectra of the northern and southern lobes; the
northern `arm' of the northern lobe is quite clearly detected at our
observing frequency, while there is absolutely no detection of any
extended emission at 108 GHz south of the eastern hotspot. The
spectral difference extends back down to GHz frequencies, in spite of
the fact that at 1.4 GHz the northern and southern lobe regions are
morphologically quite similar and have similar surface brightness. We
have not been able to rule out particle (re)acceleration at the bright
`northwest corner' of the northern lobe, which might account for the
difference, but we note that there is some detected extended emission
at 108 GHz in the northern lobe between the western hotspot and the
`northwest corner', which is not consistent with such a picture. The
difference could be caused simply by different aging
processes or different magnetic field strengths in the two regions.
However, it is tempting to relate the differences in northern and southern
lobes with the differences in the corresponding hotspots.  Specifically,
we suggest, as in the models of \markcite{meis}Meisenheimer \etal\ (1989),
that the `high-loss' eastern hotspot does not efficiently accelerate
particles to the high energies required to produce 108-GHz emission from
the lobes, while the less spectacular western hotspot is more efficient
at putting the energy supplied by the jet into high-energy electrons.

\acknowledgments
We thank the Hat Creek staff for their efforts in the construction and
operation of the long baseline array.  We would also like to thank
Matt Lehnert and Christian Kaiser for discussions, and Robert Laing
for allowing us to use his archival VLA data.  This work was supported
by NSF grants NSF-FD93-20238, NSF-FD96-13716, and AST-9314847, and
PPARC grant GR/K98582.  The National Radio Astronomy Observatory Very
Large Array is a facility of the National Science Foundation operated
under cooperative agreement by Associated Universities, Inc.

\newpage

\clearpage

\begin{deluxetable}{lrrrrrl}
\tablewidth{0pt}
\tablecaption{Flux Densities of the 3C\,123 Components\label{fluxes}}
\tablehead{\colhead{Component}&\multicolumn{5}{c}{Flux density (mJy)}&\colhead{Comment}\\
\colhead{}&\colhead{1.413 GHz}&\colhead{4.885 GHz}&\colhead{8.440 GHz}&\colhead{14.97 GHz}&\colhead{107.75 GHz}
}
\startdata
Core&$64 \pm 5$&$93 \pm 6$&$90 \pm 4$&$96 \pm 4$&$42 \pm 6$&Gaussian fit\nl
E hotspot&$15753 \pm 21$&$7200 \pm 7$&$4683 \pm 4$&$2586 \pm 4$&$303 \pm 8$\nl
\hskip 10pt(SE/E4)&\nodata&$5176 \pm 3.5$&$3519 \pm 2.1$&$1966 \pm 1.6$&$231
\pm 14$&$0\farcs64 \times 0\farcs40$ maps\nl
\hskip 10pt(NW/E3)&\nodata&$1196 \pm 2.8$&$758 \pm 1.7$&$426 \pm 1.3$&$57 \pm 11$&$0\farcs64 \times 0\farcs40$ maps\nl
W hotspot&$867 \pm 13$&$345 \pm 4$&$222 \pm 2$&$112 \pm 3$&$18 \pm 5$\nl
NW corner&$2430 \pm 12$&$911 \pm 4$&$578 \pm 3$&$302 \pm 3$&$30 \pm 5$\nl
N lobe&$11331 \pm 41$&$3136 \pm 14$&$1741 \pm 8$&$810 \pm 9$&$67 \pm 16$\nl
S lobe&$11643 \pm 45$&$2646 \pm 15$&$1179 \pm 8$&$473 \pm 10$&$<50$
\enddata
\tablecomments{Errors quoted are $1\sigma$ statistical errors based on
the r.m.s.\ off-source noise, and do not include the uncertainties in
absolute flux calibration. The upper limit is at the $3\sigma$
level. All flux densities were measured from fixed regions of matched
$3\arcsec$-resolution maps, except where specified in the `Comment'
column: see the text for details.}
\end{deluxetable}

\begin{deluxetable}{lrrrr}
\tablewidth{0pt}
\tablecaption{Two-Point Spectral Indices for the 3C\,123 Components\label{spices}}
\tablehead{\colhead{Component}&\multicolumn{4}{c}{Spectral index of Freq. Band}\\
\colhead{}&\colhead{1.4--4.9 GHz}&\colhead{4.9--8.4 GHz}&\colhead{8.4--15.0 GHz}&\colhead{15.0--107.75 GHz}
}
\startdata
Core&$-0.30 \pm 0.08$&$0.06 \pm 0.14$&$-0.11 \pm 0.11$&$0.42 \pm 0.08$\nl
E hotspot&$0.631 \pm 0.001$&$0.787 \pm 0.002$&$1.036 \pm 0.003$&$1.09 \pm 0.01$\nl
\hskip 10pt(SE/E4)&\nodata&$0.834 \pm 0.007$&$1.006 \pm 0.007$&$1.02 \pm 0.10$\nl
\hskip 10pt(NW/E3)&\nodata&$0.706 \pm 0.002$&$1.017 \pm 0.002$&$1.08 \pm 0.03$\nl
W hotspot&$0.74 \pm 0.02$&$0.81 \pm 0.03$&$1.19 \pm 0.05$&$0.93 \pm 0.15$\nl
NW corner&$0.791 \pm 0.005$&$0.83 \pm 0.01$&$1.13 \pm 0.02$&$1.17 \pm 0.09$\nl
N lobe&$1.034 \pm 0.005$&$1.08 \pm 0.01$&$1.34 \pm 0.02$&$1.26 \pm 0.12$\nl
S lobe&$1.194 \pm 0.006$&$1.48 \pm 0.02$&$1.59 \pm 0.04$&$<1.14$
\enddata
\tablecomments{
Errors quoted are derived from the $1\sigma$ statistical errors of
Table \ref{fluxes}. Spectral index $\alpha$ is defined in the sense
that flux is proportional to $\nu^{-\alpha}$.}
\end{deluxetable}

\begin{deluxetable}{lllrlrrr}
\tablewidth{0pt}
\tablecaption{Results of Synchrotron Spectral Fitting\label{fits}}
\tablehead{\colhead{Region}&\colhead{Geometry}&\colhead{Size}&\colhead{B Field}&\colhead{Best Fit}&\colhead{$E_{\rm break}$}&\colhead{$E_{\rm cutoff}$}&\colhead{$\chi^2$/dof}\\
\colhead{}&\colhead{}&\colhead{(kpc)}&\colhead{(nT)}&\colhead{model}&\colhead{(GeV)}&\colhead{(GeV)}}
\startdata
E hotspot (E3)&Cylinder&$3.5 \times 0.7$&29&Break&1.1&\nodata&1.5/2\nl
E hotspot (E4)&Cylinder&$5.4 \times 2.5$&18&Break&1.7&\nodata&11/2\nl
W hotspot&Cylinder&$4.74 \times 0.6$&20&Break&1.4&\nodata&30/3\nl
NW corner&Sphere&5&7.5&Both&1.9&15&3.4/2\nl
N lobe&Cylinder&$66 \times 12$&4.7&Both&0.55&7.9&14/2\nl
S lobe&Cylinder&$95 \times 12$&4.9&Both&0.26&4.5&23/2
\enddata
\tablecomments{
Either a spherical or cylindrical geometry was used for field-strength
fitting; the geometry chosen is given in column 2. Dimensions in
column 3 were estimated from high-resolution radio images. For
cylinders, the dimensions listed are the length $\times$ the radius;
for spheres, the dimension is the radius. Cylinders are assumed to be in the
plane of the sky. The best-fitting models in column 4 are either `Break'
(effectively, no high-energy cutoff is needed to fit the data) or
`Both' (both a spectral break and a high-energy cutoff are needed).
The electron energies for the break and, where used, the cutoff are
tabulated in columns 5 and 6.}
\end{deluxetable}

\end{document}